# CONTINUOUS IN-SITU AND REMOTE SUN OBSERVATION FOR SPACE WEATHER MONITORING AND MITIGATION OF INFRASTRUCTURE THREATS THROUGH AN OPTIMIZED HELIOCENTRIC SATELLITE CONSTELLATION

**Leonidas Askianakis**

*Technical University of Munich, Germany, Email: leonidas.askianakis(at)tum.de*

## ABSTRACT

Although vital for life on Earth, solar activity poses questions and increasing threats to humanity due to the Sun's unknown dynamics, intensified by our dependence on terrestrial and space-based infrastructure. This situation is compounded by significant gaps in our understanding of space weather phenomena, the Sun's magnetic field, and the need for rapid responses to unpredicted solar events. To address these issues, an optimized heliocentric satellite constellation is proposed that leverages satellites in an Elliptical Walker Constellation. This system offers – among others - equally distributed arguments of periapsis separations and cross-coupled true anomalies with respect to the Sun-centric coordinate frame. In this paper it is also demonstrated that this strategic multi-spacecraft configuration makes it possible to distinguish spatial and temporal changes in solar wind phenomena, reconstruct, in 3D, Coronal Mass Ejections (CMEs), predict which space or ground-based infrastructure and when it will be affected by CMEs, maintain continuous coverage of the critical Sun-Earth line throughout the mission's duration, and protect future missions by providing simultaneously in-situ and remote measurements from small and cost-effective satellites.

## 1. INTRODUCTION

Solar activity, characterized by phenomena such as sunspots, solar flares, and Coronal Mass Ejections (CMEs), has profound effects on space weather, posing significant risks to Earth's technological infrastructure. These effects include data compromise, radio interference, premature satellite deorbit, power grid failures, and GNSS data compromise. The unpredictability, intensity, and significance of these solar events necessitate continuous and comprehensive monitoring of the Sun [17], [18], [19], [20].

Despite substantial advancements through missions like SOHO, STEREO, and the Parker Solar Probe, persistent gaps still need to be addressed, particularly in achieving uninterrupted solar monitoring for the entire Sun, accurate predictions of solar events, and mitigation techniques in case of unpredicted scenarios. Existing missions are limited by factors such as constrained observational latitudes, lack of continuous Sun-Earth line coverage, and absence of in-situ measurements. These limitations impede the ability to effectively predict and respond to solar weather events.

The proposed mission aims to address these challenges through an optimized heliocentric satellite constellation, utilizing an Elliptical Walker Constellation. This configuration ensures continuous $4\pi$-steradian coverage of the Sun's surface, with a minimum 10% overlap in coverage. The constellation is designed to meet multiple objectives simultaneously, offering comprehensive spatial and temporal analysis of solar phenomena, 3D reconstruction of CMEs, and continuous Sun-Earth line monitoring.

The uniqueness of this constellation lies in its ability to provide both in-situ and remote measurements simultaneously. This multi-objective approach not only enhances scientific understanding but also reduces costs compared to conducting multiple separate missions.

## 2. MISSION OVERVIEW

### 2.1. Existing Missions

While analyzing the capabilities and objectives of existing and upcoming missions, multiple shortcomings have been identified interchangeably between them. A summary of those follows in the following table, and a detailed assignment of each of those to the relevant existing mission follows in the appendix.

| | |
|---|---|
| (a) | Only constrained and non-continuous observations of the Sun’s poles |
| (b) | No/limited stereoscopic analysis and 3D reconstruction capabilities |
| (c) | No 3D localization of radio bursts |
| (d) | Limited and discontinuous forecasting of the arrival of CMEs and Solar Energetic Particles (SEP) |
| (e) | No continuous monitoring of the Sun-Earth Line (SEL) |
| (f) | No high-resolution imaging of sunspots and CMEs |
| (g) | Does not provide in-situ measurements |
| (h) | Outdated and faulty instruments |

Table 1. Summary of existing mission limitations

During the initial stages of Phase 0 of the study, the analysis led us to identify critical areas where improvements are necessary to ensure the safety of both humans and infrastructure on and around Earth. Further, an analysis of the state of the art was followed to identify existing architectures and mission concepts that attempt to solve individual or multiple of these limitations simultaneously.

### 2.2. Mission Objectives

With this analysis in mind, the following mission objectives have been identified and analyzed further during the design of the mission. Both the state of the art in observation missions around the Sun and the capabilities of different instruments led to their creation.

| Mission Objective (M) | Description |
|---|---|
| M1 | Continuous 4π-steradian (sr) observation of the Sun's surface with at least 10% overlap |
| M2 | Monitoring Coronal Mass Ejections (CMEs) and sunspots with high-resolution magnetic data |
| M3 | Providing observation windows for 3D reconstruction of CMEs between the satellites |
| M4 | Constant monitoring of the Sun-Earth Line (SEL) is constrained in a cone of observation with an angle of 10 degrees. |
| M5 | Detecting and tracking Type-II radio bursts |
| M6 | Detecting and tracking Type-III radio bursts |
| M7 | Localizing in 3D Type-II radio bursts |
| M8 | Localizing in 3D Type-III radio bursts |
| M9 | Forecasting the arrival of CMEs associated with Type-II radio bursts at Earth. |
| M10 | Forecasting the arrival of SEP electrons associated with Type-III radio bursts at Earth. |
| M11 | Forecasting the potential arrival of SEP protons associated with Type-III radio bursts |
| M12 | Providing in-situ measurements of the inner heliosphere (0.1-0.3 AU) |
| M13 | Providing observation windows for 3D stereoscopic analysis of magnetic loop geometries of with two satellites forming an angle of 10 degrees or less on the Satellite-Sun-Satellite (SSS) triangle. |

Table 2. Mission Objectives

## 3. SCIENTIFIC ANALYSIS

### 3.1. Continuous 4π-Steradian Coverage

The full 4π-observation coverage of the Sun's heliosphere is critical for a comprehensive understanding of solar phenomena and their impact on space weather. This coverage ensures that all regions of the Sun are observed simultaneously, allowing for continuous monitoring of solar activities, solar flares, and solar wind interactions with the interplanetary medium.

By observing the poles and the equatorial regions simultaneously, we can understand holistically the creation of solar events and how these are potentially coupled with effects and conditions across different regions of the Sun [24]. On top of that, for accurate space weather prediction, real-time data from multiple observation points around the Sun have become necessary to achieve higher accuracy.

Full sphere observations will help uncover the heliosphere's structure, the solar wind's distribution, and the overall heliospheric environment [26]. These observations will also allow the identification of the origin of the slow solar wind streams. A source that still needs to be better understood [27].

Current missions like SOHO, STEREO, and Solar Orbiter have significantly contributed to our understanding of solar dynamics. However, they are limited by their observational geometry and temporal coverage of specific regions of the Sun. To further increase our holistic understanding of the Sun's dynamics, simultaneous coverage of the entire heliosphere has proven necessary.

### 3.2. In-situ Measurements

The observation of the Sun's inner heliosphere lacks data in crucial regions like the one of the poles and, at the same time, is constrained. The origin of this problem is directly connected to the high eccentricities of the existing solar orbiters and the proximity encounters, which enable only short in-situ observation intervals over a mission's lifetime. Therefore, the need for available data on the Sun's inner heliosphere over various time scales and observation points is increasing. This need led us to require the mission to visit the inner heliosphere, for which an instrumentation and objective analysis have been performed. Specifically, a required distance of 0.1-0.3AU has been identified so that gamma-ray and X-ray measurements, along with energetic particle detectors like EPD [34] and magnetometers like (MAG) [35].

### 3.3. Coronal Mass Ejections Analysis and Sun-Earth Line Monitoring

CMEs are among the most significant solar events impacting space weather, with the potential to cause severe disturbances to Earth's technological infrastructure [18]. Accurate and timely detection of CMEs, particularly those directed along the Sun-Earth line (SEL), which would imminently influence Earth, is crucial for mitigating their adverse effects. Firstly, CMEs are primary drivers of geomagnetic storms, and early detection of them allows for implementing preventive measures, such as adjusting satellite orbits and managing power loads in grids, to mitigate these disruptions [28].

Moreover, continuous monitoring of the SEL ensures that any unpredicted solar events, such as radio bursts and fast-evolving CMEs, are promptly detected. This capability is essential for timely warnings and rapid response to potential threats.

The deployment of spacecraft equipped with instruments capable of near real-time CME detection and radio burst monitoring placed on the SEL can significantly improve our readiness for solar events. Instruments such as coronagraphs and radio spectrographs, integrated with automated detection algorithms, can promptly identify CMEs and their associated phenomena and issue warnings that would enable mitigation of their effects. An example of similar architecture, with limited observation scope and capabilities, is the Solar Orbiter's METIS coronagraph, which employs on-board algorithms to detect CMEs and assess their propagation towards Earth [29].

Furthermore, integrating upstream measurements of radio bursts and solar winds enhances the characterization of solar phenomena. Observations of type II and III radio bursts, which often accompany CMEs, provide insights into the acceleration of solar energetic particles and the dynamics of shock waves in the corona [31]. When combined with solar wind data, these measurements offer a comprehensive view of the solar environment and its potential impacts on Earth, and when this takes place between the Sun and Earth, the effectivity of our Space Weather mitigation techniques is tremendously increased.

The ability to detect and monitor CMEs and other solar events in the SEL provides significant safety and mitigation benefits. Timely warnings enable operators of critical infrastructure to implement protective measures, reducing the risk of damage and ensuring the continued operation of essential services. Additionally, accurate predictions of solar events enhance the reliability of space weather forecasts, contributing, among others, to the safety of astronauts and spacecraft in orbit. It has also been estimated that mitigating the effects of a single unpredictable event, like the 2003 Halloween Storm (with an occurrence likelihood of once every 20 years), can benefit the global economy for at least 27 billion USD worth of infrastructure damage [32].

### 3.4. 3D Geometrical and Magnetic Stereoscopy of Coronal Structures

Understanding the complex dynamics of the Sun's atmosphere and its influence on space weather requires detailed 3D and magnetic stereoscopic analysis of coronal structures. This is especially critical in regions where CMEs and solar flares originate, as these phenomena significantly influence the further propagation of the flares.

The capability for magnetic stereoscopy and the stereoscopic reconstruction of magnetic loops at CME and flare origin sites is currently inadequate in terms of both quality and quantity [13]. This limitation hampers our understanding of the initiation and development of CMEs, which are primary drivers of space weather events. Improved magnetic loop data would enhance our predictive models of these solar events, providing more accurate forecasts and mitigating potential risks to Earth-based and space-based systems [1].

To fully understand the interaction of CMEs with the interstellar medium, it is necessary to achieve precise 3D geometric reconstructions of CMEs as they propagate through space [14]. Such reconstructions allow us to visualize the structure and trajectory of CMEs, leading to better predictions of their impact on Earth. The inability of the STEREO mission to extract detailed magnetic loop geometries due to the poor image quality of the EUVI instrument underscores the need for state-of-the-art spacecraft equipped with high-resolution instruments [1].

The proposed mission design overcomes these limitations by employing advanced instruments with higher angular resolution than those used in current missions like STEREO. For instance, replacing the Atmospheric Imaging Assembly (AIA) with newer, more capable instruments will increase the allowable separation between spacecraft for magnetic loop reconstruction, estimated to be within a 10-degree separation during observation windows. This improvement will enable more accurate and detailed analyses of the Sun's magnetic field dynamics while relaxing the relevant requirements that enable these observations.

## 4. ASTRODYNAMICS

### 4.1. Orbital Configuration Design

To determine the optimal orbital configuration of the constellation, we started by studying the extensive list of mission objectives and grouping the orbital features of the spacecraft with which objectives they can achieve simultaneously.

The resulting concept was an Elliptical Walker-like constellation [5], on which eccentric orbits with the Sun on the one loci and rotary symmetric increase of the arguments of periapsis (as detailed below). The concept we concluded could conceptually achieve all the objectives simultaneously, and the detailed design remained. For this, we delve deeper into the calculations below on which the orbital elements refer to the heliocentric ones.

### Inclinations assignment

From objective M1 already, each satellite's orbital plane's inclination is fixed to achieve the desired measurements in the Sun's polar regions. These, together with considerations of the zenith viewing angles of the polar regions and the considerations of the Δv requirements and the Venus fly-bys (as detailed below), provided the final inclination value.

The minimization of the zenith viewing (angle between $\vec{r}_p$ and $\vec{r}_n$ as shown in Figure 1 below) angle itself is beneficial for the quality of the data gathered from the spacecraft since, for smaller angles, the light and particles originating from the poles travel less through the Sun's heliosphere and the interstellar medium. This behavior is described by Beer's law [33]

$$I = I_o e^{-\frac{x}{\lambda}} \quad (1)$$

which was considered during the optimization process. Specifically, the traveling distance of light through the interstellar medium can be calculated based on the zenith viewing angle, and therefore a direct relation for the optimization process was directly identified.

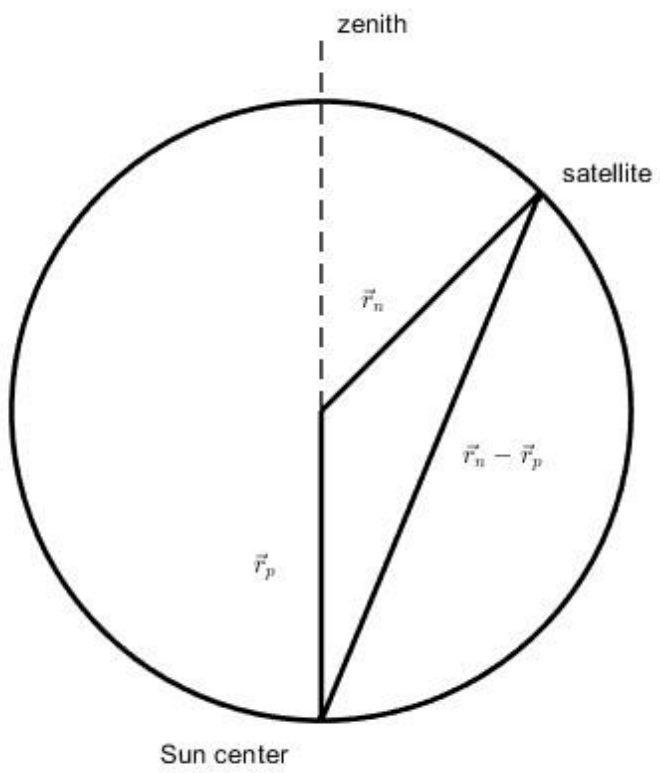

Figure 1. Zenith viewing angle

Alternative concepts that could be analyzed in the future include the further optimization of the inclination assignments. This assignment could instead happen pair-symmetric, or in general on repetitive patterns of different frequencies, such that only a fraction of the satellites of the total constellation is used to study the polar regions constantly, without the need for every satellite to have a high inclination, that would cost more Δv to achieve the orbits. The synchronization of their arrival at the target locations on the ecliptic plane and in the vicinity of the SEL would be mandatory and could be achieved by further optimizing the true anomalies of such orbits, an example of which is given below.

### RAAN separations assignment

For symmetry reasons and to provide constant and controllable properties between the spacecraft, as it has been explained so far, the assignment of RAAN separations follows a rotary symmetric configuration. This is directly related to the amount of spacecraft by:

$$RAAN_i = \frac{360}{n}\,(i-1)\ ,\, i \in \{1, 2, \dots, n\} \quad (2)$$

and their distribution directly influences the orbital period that we assigned to each satellite due to the constant SEL requirement and, therefore, maximization of the time that each spacecraft spends inside the CME dispersion cone of interest.

It is visible that the RAAN separations of the satellite orbits depend directly on the number of satellites that are included in the constellation since all of them are symmetrically separated. Later work will include the individual separations of the RAANs of every satellite's orbit, optimizing every two satellites instead of as a total. This is to allow the non-symmetrically optimized configuration, which will be optimal for the specific date of launch of the mission, by purposefully concentrating more satellite coverage in one part of the 360 degrees of the heliocentric plane, In this way, higher/denser coverages for the duration of the mission will be achieved (like for example for the SEL coverage) while keeping the same amount of spacecraft. This non-symmetric configuration would indeed offer better results for the (non-infinite) duration of such a mission, although mathematically, it would induce non-even distribution and less comprehensive coverage if the time/duration of the mission would go to infinity. Since our mission is not designed to last forever, an asymmetric separation of RAAN distributions could offer optimized coverage results according to the mission profile's needs, which will be included in future work.

### Semi-major axis and eccentricity assignment

The semi-major axis is fixed by the required period of the orbit, which is additionally required to be synchronized with the RAAN separations and the observation requirements of the mission.

The constraints for the semi-major axis originate from multiple simultaneous mission objectives. First, on the science side, a distance $d \in [0.1, 0.3]\, AU$ to the Sun is required to achieve the necessary conditions for the required in-situ measurements to take place while maintaining a distance that maintains the design of the mission feasible. Therefore:

$$0.1\,AU < r_{periapsis} < 0.3\,AU \quad (3)$$

Second, in order to ensure that each satellite is contributing to the objective of the Sun-Earth Line coverage, an additional constraint of

$$r_{apoapsis} < 1\,AU \quad (4)$$

is introduced. Finally, in order to achieve coverage of the SEL on both the apoapsis and the periapsis of each satellite's orbit – which are necessary to maximize the coverage as long as the constellation is in such a symmetrical configuration – the half-period of Earth has to be integer multiplier of the half-periods of the satellites' orbits, while synchronously arriving at their initial positions (here assumed apoapsis), and therefore

$$2\nu\frac{1}{2}T_{Earth} = 2\mu\frac{1}{2}T_{Satellite} \quad \mu,\nu \in \mathbb{N}^*$$
$$T_{Earth} = \frac{\mu}{\nu}T_{Satellite} \quad \mu,\nu \in \mathbb{N}^* \quad (5)$$

Moreover, to achieve uniform coverage of the SEL and maintain symmetric observation conditions between the satellites' and Earth's positions, an additional symmetry condition has to be established at every revolution of Earth around the Sun. This means that the time instances that the Earth and satellites reach their periapsis have to be synchronized, which results into

$$(2\lambda+1)\frac{1}{2}T_{Earth} = (2\kappa+1)\frac{1}{2}T_{Satellite} \quad \lambda,\kappa \in \mathbb{N}$$

$$T_{Earth} = \frac{(2\kappa+1)}{(2\lambda+1)}\,T_{Satellite} \quad \kappa,\lambda \in \mathbb{N} \quad (6)$$

Then, by combining Eqs. (4) and (6), and minimizing the coefficients – in order to achieve longer observation windows for both in-situ measurements and 3D stereoscopy – we end up with the final optimal coefficients that will eventually determine the orbital period, and hence the semi-major axis of the individual satellite orbits.

$$\begin{cases} T_{Earth} = \min\limits_{\kappa,\lambda}\left(\dfrac{2\kappa+1}{2\lambda+1}\right) T_{Satellite} \quad \kappa,\lambda \in \mathbb{N} \\ r_{apoapsis} < 1\,AU \end{cases} \Rightarrow$$

$$T_{Earth} = 3T_{Satellite} \quad \kappa = 1, \lambda = 0 \quad \Rightarrow$$

$$a_{Satellite} \cong 0.48\,AU \quad (7)$$

Under these circumstances and constraints, the satellite configuration is uniquely determined by these coefficients, and this exact uniqueness in the solution of the system of relations is what enabled the mission design in the first place and what mathematically ensures the optimality of the design in satisfying all the objectives at the same time.

Using the relations (3), (4), and the resulting semi-major axis of 0.48 AU, we can extract the allowed range for the eccentricities of the orbits.

$$\begin{cases} 0.1\,AU < r_{periapsis} < 0.3\,AU \\ a_{Satellite} \cong 0.48\,AU \\ r_{apoapsis} < 1\,AU \end{cases} \Rightarrow$$

$$0.1\,AU < 0.48\,AU\,(1 - e_{Satellite}) < 0.3\,AU \Rightarrow$$

$$0.37 < e_{Satellite} < 0.79 \quad (8)$$

The flexibility of this range allows the optimization of the Venus fly-bys according to the chosen epoch of the mission and the launch window.

**Arguments of periapsis assignment**

Then, the arguments for periapsis for all the satellites were determined. The different inclinations between the ecliptic plane and all the satellite orbital planes, together with the requirement of maximum SEL coverage, were considered. The arguments for the periapsis of every satellite had to be individually and independently specified to achieve the line of periapsis that lies on the ecliptic plane for every satellite. This constraint is in place to maximize the coverage that every satellite provides to the SEL mission objective, therefore keeping its closest and further approaches to the Sun (on which significant interval of the orbital duration is spent) between the Earth and Sun. Having this way achieved a mathematical condition for the optimality of the chosen arguments of periapsis (in the symmetric scenario).

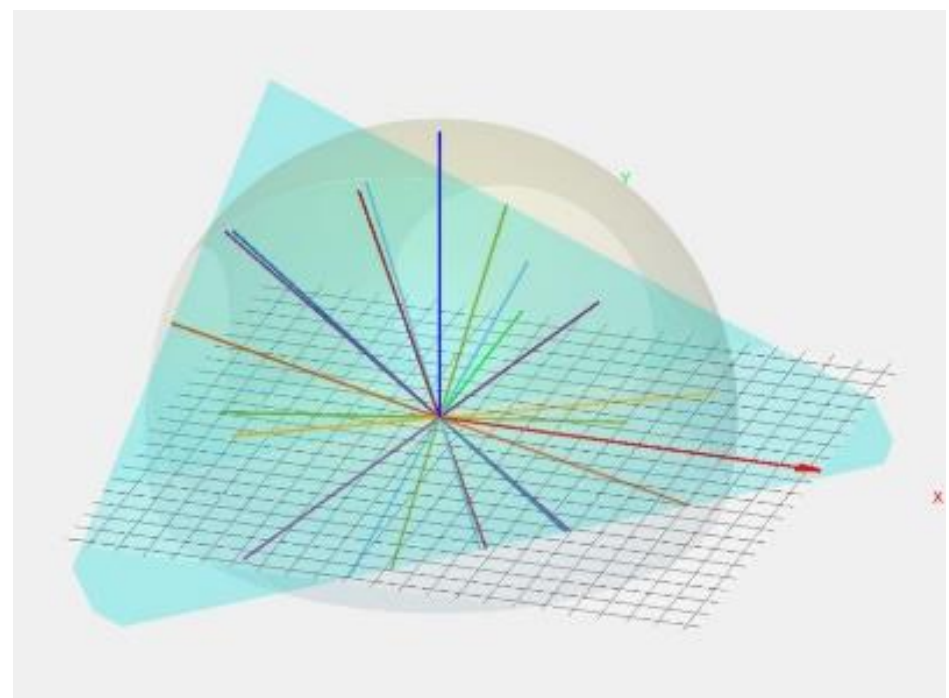

Figure 2. Numerical optimization of the argument of periapsis

Under these constraints, we performed Runge-Kutta-4-based numerical optimization to satisfy the needs of the mission, and the resulting arguments of periapsis defined the further optimization procedure. The calculation of the intersections of each satellite's orbital plane (colorful lines) with the Earth's plane (light blue) and the optimization of their separations (by controlling the argument of periapsis) is showcased in Figure 2 above. This served as the basis of the identification of the optimal arguments of periapsis to achieve even separations between the intersection of the orbits such that we ensure that the semi-major axis lies on the Earth's orbital plane.

**True anomalies optimization**

After all the previous orbital considerations, the only unconstrained orbital parameter was (fortunately) the true anomaly of every satellite and the one for which we directly optimized. The exact separation and assignments of the true anomalies directly depend on the SEL objective (M4) and the 3D stereoscopic analysis observation windows objectives (M3, M13). Therefore, the true anomalies of all the satellites are coupled with each other, constraining this way their range to the final solution of the orbital configuration. After performing this level of constrained optimization, the first converged solution is visualized in the figure 3 below.

The orbital planes of each satellite are clearly not parallel to the ecliptic; the inclinations allow for direct observation of the poles and the SEL coverage during both the encounter of the apoapsis and the periapsis of the spacecraft's orbits. The detailed results are detailed in the resulting orbital configuration section.

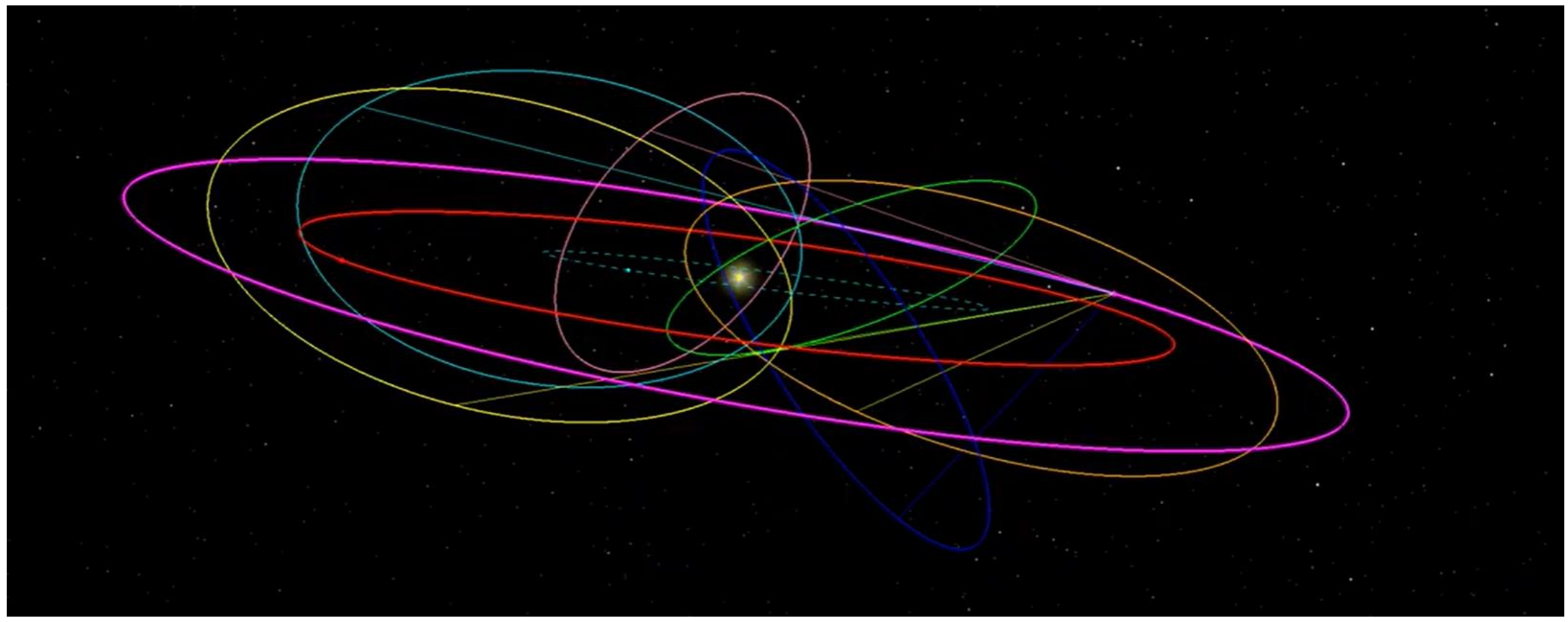

Figure 3. The first orbital configuration solution

**Optimization Summary**

The consideration/employment of every orbital parameter of a constellation and the inclusion of unique constellation designs, like the one of the Elliptical Walker constellation, on the design, allowed mathematically the solution of this constrained optimization problem. If the same optimization over these parameters was executed with a numerical brute force approach over all the different orbital parameters and constellation designs instead of analytically, there would be a need for tremendous computational resources. Specifically, if commercial tools, widely used in the domain, were employed or methods like [10], [11] were used, the computational demands would remain unfeasible for the mission design lifespan even for NASA's Pleiades supercomputers.

It is important to note that we have identified the later methods to be of particular importance for the further future optimization of the constellation design, which will include factors like sensor specificities, launch window optimization, and fly-by trajectory designs, which will be detailed later.

### 4.2. Stereoscopy Observation Windows Design

All the stereoscopic analysis methods described require different separation angles on the Satellite-Sun-Satellite triangle for optimal results. Different angles are optimal for magnetic stereoscopy of the loop geometries at the point of origin of the CMEs and different for the 3D geometric reconstruction of the CMEs themselves. The solution to this problem comes by continuously altering the separation angle between the satellites and pairwise coupling their true anomalies of the satellites in order to enable all of these objectives with the same mission design. The proposed architecture ensures observation windows on which two satellites form an angle on the SSS triangle of less than 10 degrees by coupling the true anomalies so that when a satellite approaches the apoapsis, another satellite leaves the apoapsis at close proximity. Ensuring this way a time window on which the two satellites are spaced closely apart and that both 3D reconstruction of CMEs and magnetic stereoscopy are achieved. This characteristic, in particular, uniquely allows the fulfillment of both M13 and M4 mission objectives during the passage of the two spacecraft from their respective apoapsis.

In the below example (Figures 4 and 5), the locations of the satellites with their respective Satellite-Earth line can be seen. At the given instance, satellite 4 and satellite 6 of the constellation are placed closely enough (less than 10 degrees on the SSS triangle) so that stereoscopic measurements are enabled – as seen at the snapshot point of Figure 5. The marked orange dot and marked green dot represent the spacecrafts at their close encounter with the Sun during their apoapsis phase.

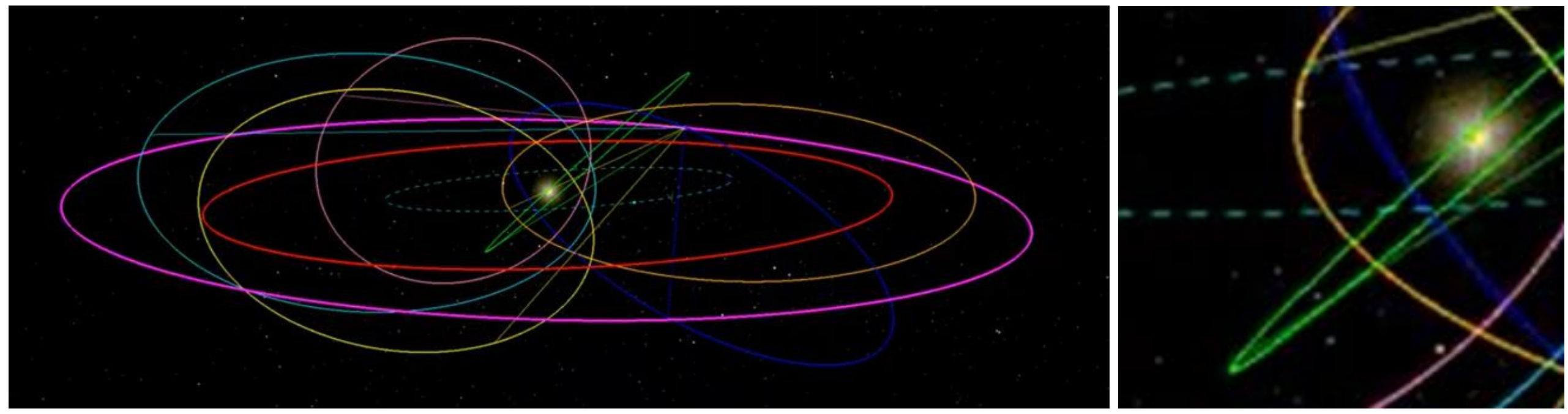

Figures 4 and 5. Close encounter of satellite 4 (orange) and 6 (green)

### 4.3. Trajectory Design

By employing methodologies like EDVEGA [3], [4] for electric propulsion, in combination with Earth and Venus gravity assists, highly elliptical with respect to the ecliptic plane orbits could be achieved at a well-decreased requirement for Δv generated by thrusters.

According to [4], around 7 km/s of Delta-v would be needed to achieve such a highly elliptical orbit in this configuration. An even better performance could be achieved by further optimizing the EDVEGA methodology with nonlinear programming. By timely offsetting the dispatches of every satellite from Earth, the different characteristics of the constellation could be achieved, reaching Venus at different time instances and manipulating both the inclination change and the Sun-centric speed of the spacecraft.

In the mission design of SUNFLOWER [7], it has been shown that 29 km/s of Δv can be achieved using the Next-C gridded ion engine, with a total of more than 38 km/s Δv artificially achieved without any gravity assist, and still classifying and proving the mission as feasible. At the same time, Solar Orbiter will achieve a 33-degree inclination with just a fraction of the previously proposed Δv, thanks to the Venus fly-bys. In a similar fashion, our proposed design will take advantage of multiple - timely offset - gravity-assists on Venus to achieve the final configuration with higher Δv efficiency.

Considering the pre-mentioned considerations, our initial calculations provide estimates of Δv requirements in the range of 19 km/s per satellite, a metric that proves it by far feasible according to [7].

### 4.4. Resulting Orbital Configuration

The resulting orbital configuration includes 6 spacecraft with equally separated RAANs at an eccentricity of 0.65, a semi-major axis of 0.48 AU, an inclination of 47 degrees, and true anomalies spaced such that the necessary objectives are fulfilled, and the arguments of periapsis that achieve the requirement of the apo-peri apsis line to lie on the equatorial plane.

The Satellite-Earth lines are also visualized in all the previous figures to showcase the placement of each satellite relative to Earth at any given point. Something that was then used for the communications disruption analysis from the Sun, as detailed below. Below it can be seen in figure 6 the final configuration of the converged simulation, with all the satellites out of the ecliptic, and evenly spaced RAAN distributions. A more detailed overview can be seen in the following link, on which a snippet of the simulation is provided.

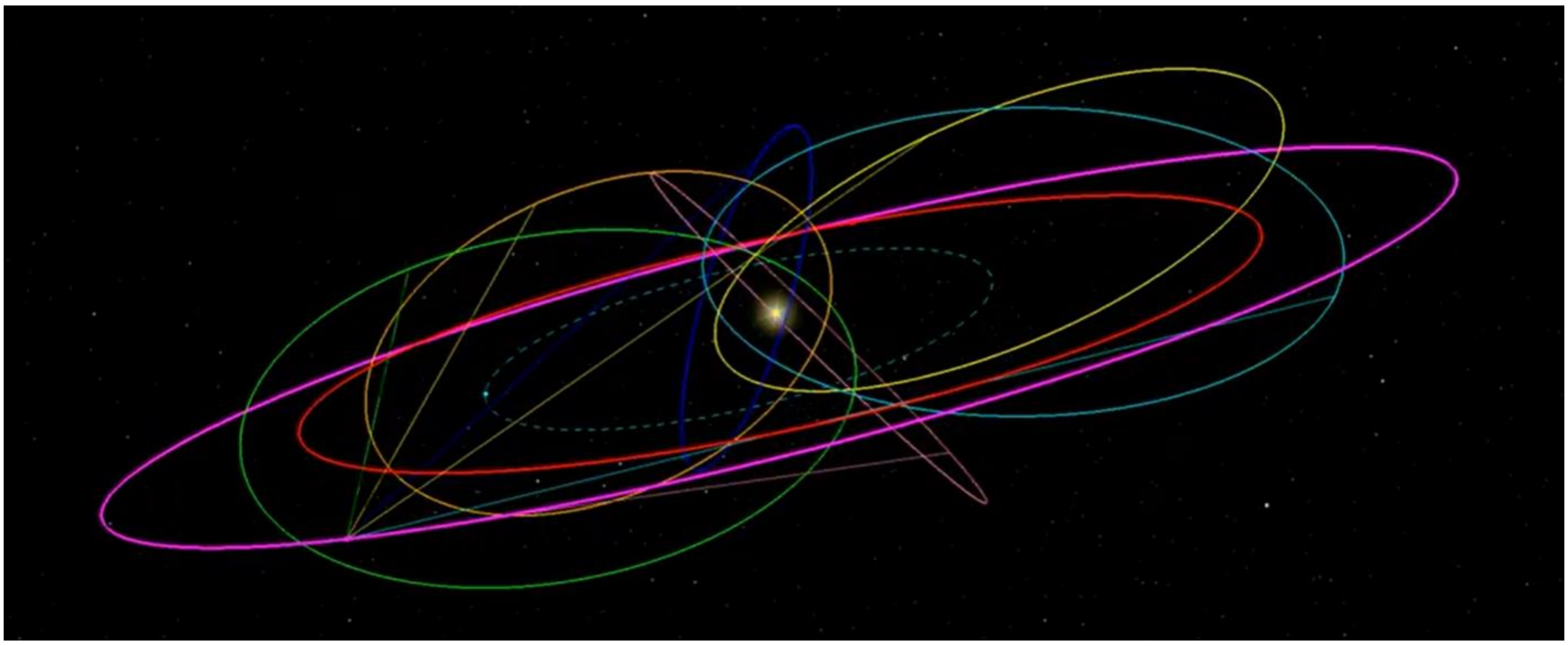

Figure 6. The final constellation configuration

## 5. COMMUNICATIONS

For the given satellite configuration, a telecommunications disruption analysis has also been performed to validate the healthy communication link between satellites and Earth for most of the orbital period of each satellite. For this, the Satellite-Earth-Satellite (SES) angle analysis on the respective triangle has been analyzed.

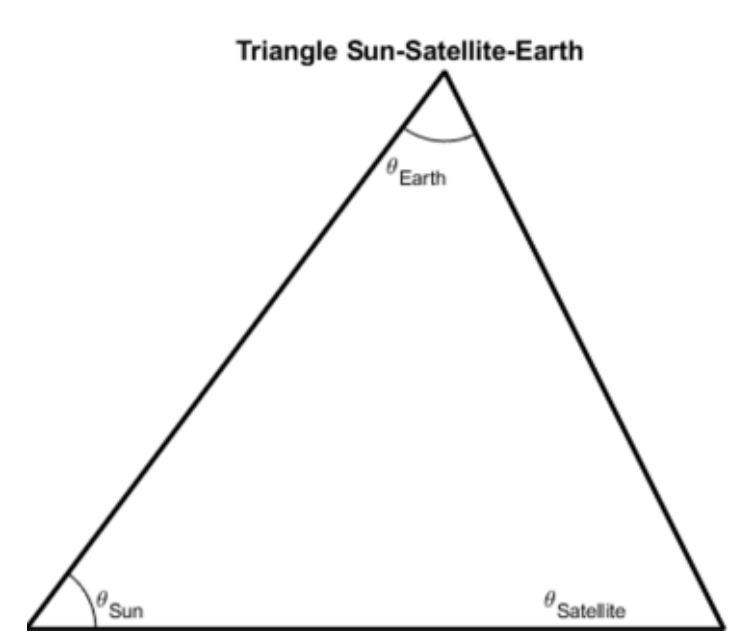

Figure 7. Sun-Earth-Satellite triangle

The goal of this analysis was to demonstrate that a high-reliability communications link can be achieved and that the close overlap of the Sun's magnetic field and the EM pulses during CME events will not prevent the acquisition of the mission's data or fulfill the mission objectives. At the same time, the disruption regions for each satellite have been identified, leading this way to a proactive communications sub-system design. Due to the highly elliptical shape of the constellation design, we can achieve relatively small fractions of such disruption windows, managing this way to have a less constraint communication system than the one of the Parker Solar Probe or the Solar Orbiter that also achieves in-situ measurements.

In Figure 9, the magnitude of the relative angles of the satellites for the duration of one orbit is visualized. The communication disruptions occur only once the angles are below 10 degrees, and every second, a narrow "double jump" region, on which the reference for the angle is swapped (we plot the magnitude), followed by an interval of high relative angle between the spacecraft and Earth. This duration of approximately 6% of the total orbital period is the only period that communication with a satellite is lost (once on the opposite side of the Sun) due to the synchronization of the highly elliptical orbits. In this way, almost constant communication and live updates for the entire sphere of the Sun are achieved, achieving an additional milestone. If inter-satellite links were to be implemented (complexifying this way the design), a constant $4.4\pi$ coverage could be achieved.

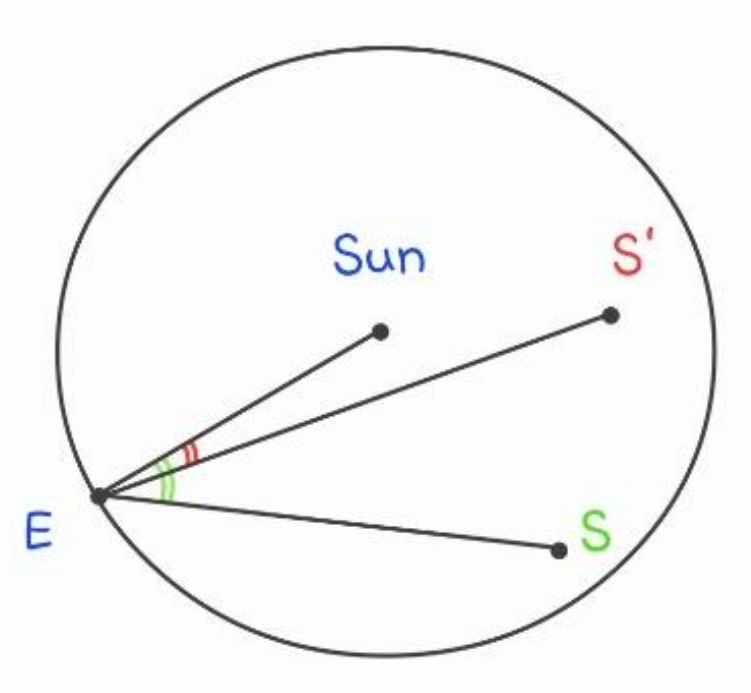

Figure 8. Inscribed angles of S' and S spacecraft

The non-regularities appearing on the patterns between the angles are caused by the simultaneous relative motion of Earth around the Sun. It can also be seen that some graphs show symmetry over the y-axis by two, indicating the offset between a satellite "preceding" or "succeeding" Earth while fulfilling the SEL mission objective (M4). The curves would have a theoretical upper limit of less than 90 degrees. This is the case because the satellite will always be inside the elliptical disk defined by Earth's orbit, and the SES angle is inscribed on the circumference of it, as shown in Figure 8. In order to optimize the percentage of a satellite's orbit on which objective M4 is fulfilled, they are purposefully kept into range. The further optimization of the duration of the orbits of each satellite between Earth and the Sun, "guarding" the SEL, is part of future work.

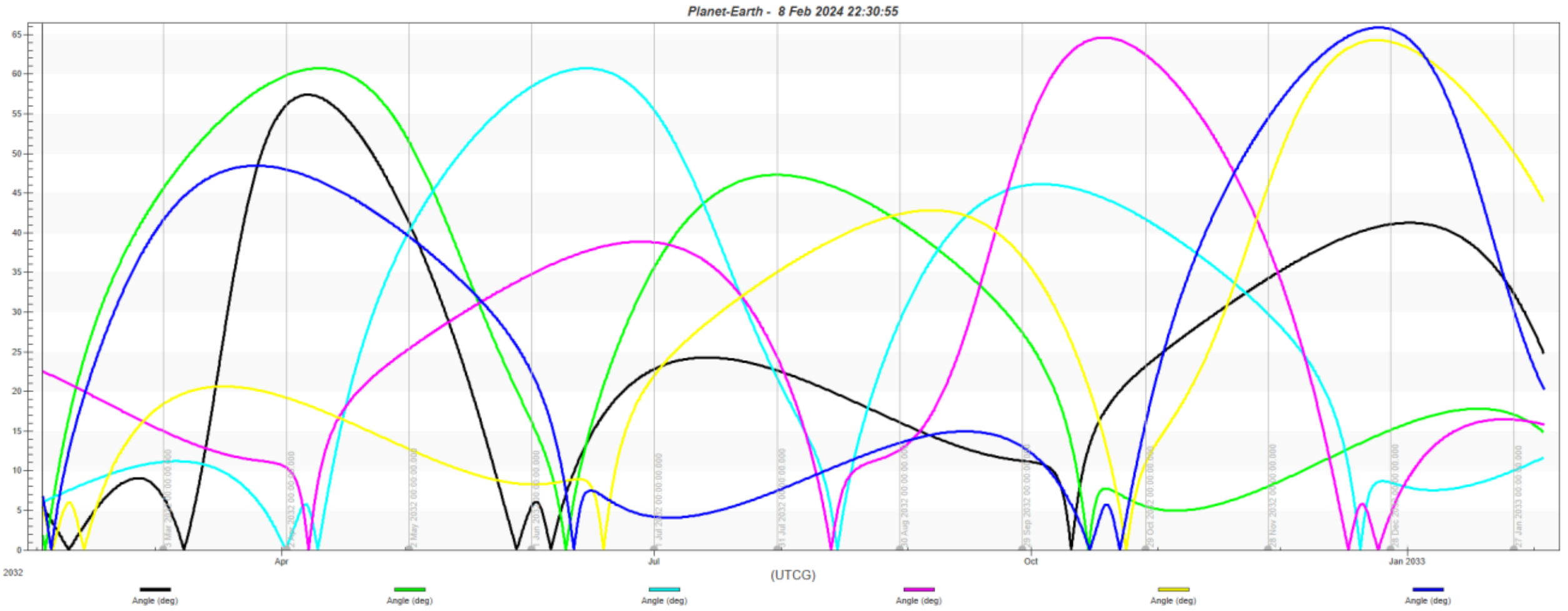

Figure 8. SES-angle analysis

## 6. FUTURE WORK

Although the design has been carried out to its fullest extent and an optimal (symmetric) orbital configuration for the mission has been identified, further improvement could be inherited in some specific categories. Specifically, further improvement on the orbit insertion maneuvers and the synchronization with the launch date and the respective eccentricities will be necessary. Additionally, the concepts mentioned above of non-symmetric assignment of RAANs and arguments of periapsis for each satellite could be inherited, together with the inclination optimization for the satellites over pairs. This would allow further improvements from a cost point of view for the mission, but it is important to note that the exact solution to this optimization problem is not guaranteed, as the system will be overconstrained, and further methodologies for the optimization will be necessary. To this end, the finalization of the non-dominated sorting genetic algorithm III (NSGA-III) [10][11] for the constellation design and fine-tuning, which was already partially implemented for the current design, will be necessary. The goal will be to optimize simultaneously for all the constraints and tradeoffs of the Heliocentric Constellation, both from a scientific perspective and a cost and feasibility analysis, like the selection of the thruster, the development timeline, and the inter-satellite links. Also, the study of launcher packing and unpacking will be studied, for which advanced robotic techniques [36] could be leveraged to ensure that higher filling factors of the launcher fairings could be practically unpacked and utilized – fitting this way larger structural assemblies onboard.

Further, the thermal, telecommunications, and launcher analysis has already been conducted on the mission, and the rationale behind those will be detailed in future work. Also, the detailed trajectory design to achieve with minimal $\Delta v$ effort the final orbits of each satellite and the necessary time-offset fly-bys have also to be given in future work, for which the detailed characteristics of the thrusters selected will be necessary.

In the end, if the mission objectives require it, and the budget of such a mission allows it, there could be a further optimization of the couplings between true anomalies. This would consider the observation windows that enable extra observation objectives for the constellation with output characteristics like the ones of Helio-Swarm, on which multi-scale tetrahedra formations are required. Such an objective could be satisfied by coupling the true anomalies so that observation windows exist, on which 4 satellites meet together at a more limited duration and frequency during the lifetime, either on top of the polar or other Sun regions.

## 7. CONCLUSION

Based on the comprehensive analysis showcased in this paper, we identified a mission design capable of achieving simultaneously the most scientific and civil-related objectives than any other existing mission. Many of the newly introduced objectives, like the continuous coverage of the safety-critical SEL (M4, M9, M10, M11) and the achievement of the implementation of emergency response mechanisms on Earth against solar storms, would help humanity save billions worth of potential infrastructure damage – accomplishments that no existing design can achieve. At the same time, the inclusion of diverse objectives and scientific capabilities, like both the stereoscopic analysis of critical points around the Sun (M3, M13) and the unprecedented full-sphere optical and magnetic coverage (M1, M2), will provide a tremendous amount and quality of data to contribute on our quest to understand our home star. All these, while combining the in-situ measurements capabilities (M12) that other multi-billion dollar missions, like the ones of Parker Solar Probe and Solar Orbiter, can uniquely achieve so far.

In detail, by employing an Elliptical Walker-like constellation, this mission ensures continuous 4π-steradian coverage of the Sun's surface, with a focus on monitoring coronal mass ejections (CMEs) and providing real-time 3D reconstructions of solar phenomena. The unique orbital configuration facilitates constant monitoring of the Sun-Earth line, enhancing our ability to predict and respond to geomagnetic storms and other space weather events.

The integration of advanced instruments capable of high-resolution magnetic data acquisition and the ability to detect type-II and III radio bursts (M5, M6, M7, M8) allows for a comprehensive understanding of solar dynamics. This approach not only enables accurate forecasting of CME impacts on Earth but also enhances our readiness for unforeseen solar events, thereby safeguarding technological infrastructure and contributing to global space weather resilience.

In addition, the design is directly re-producible in case new mission objectives or requirements due to future payload are integrated, making this way the design mathematically optimal for all the fine tunings that might be required during the upcoming phases of the study and, if adopted, even a PDR.

Overall, the fact that a more complex design view was inherited allows the inclusion of more optimization parameters for the entire constellation, a feature that will enable further fine-tuning of the design parameters according to the development process and needs. At the same time, modularity can be ensured by providing the ability to increase the observation intervals or the redundancy of the measurements by integrating further spacecraft - even during the lifetime of the mission - and constraining the optimization accordingly.

## 8. ACKNOWLEDGEMENTS

We would like to acknowledge the pre-print status of this manuscript, with it being already submitted for a potential publication at the Journal of Space Safety Engineering.

## 10. APPENDIX

| Solar Orbiter [8] | - (a) Only constrained and non-continuous observations of the Sun's poles<br>- (b) Limited stereoscopic analysis and 3D reconstruction capabilities<br>- (c) No 3D localization of radio bursts<br>- (d) Limited and discontinuous forecasting of the arrival of CMEs and Solar Energetic Particles (SEP)<br>- (e) No continuous monitoring of the Sun-Earth Line (SEL) |
|---|---|
| Parker Solar Probe [21] | - (a.2) Incapability to study the poles of the Sun<br>- (b.2) No stereoscopic analysis and 3D reconstruction capabilities<br>- (c) No 3D localization of radio bursts<br>- (d) Limited and discontinuous forecasting of the arrival of CMEs and Solar Energetic Particles (SEP)<br>- (e) No continuous monitoring of the SEL |
| Ulysses [22] | - (a) Only constrained and non-continuous observations of the Sun's poles<br>- (b.2) No stereoscopic analysis and 3D reconstruction capabilities<br>- (c) No 3D localization of radio bursts<br>- (d) Limited and discontinuous forecasting of the arrival of CMEs and Solar Energetic Particles (SEP)<br>- (e) No continuous monitoring of the Sun-Earth Line (SEL)<br>- (f) No high-resolution imaging of sunspots and CMEs<br>- (g) Does not provide in-situ measurements |
| STEREO [1] | - (a.2) Incapability to study the poles of the Sun<br>- (d) Limited and discontinuous forecasting of the arrival of CMEs and Solar Energetic Particles (SEP)<br>- (e) No continuous monitoring of the Sun-Earth Line (SEL)<br>- (f) No high-resolution imaging of sunspots and CMEs<br>- (g) Does not provide in-situ measurements<br>- (h) Poor image quality of the EUVI instrument, limited extraction of magnetic loop geometries |

| SOHO [23] | - (a.2) Incapability to study the poles of the Sun<br>- (b.2) No stereoscopic analysis and 3D reconstruction capabilities<br>- (c) No 3D localization of radio bursts<br>- (f) No high-resolution imaging of sunspots and CMEs<br>- (g) Does not provide in-situ measurements |
|---|---|
| SDO [24] | - (a.2) Incapability to study the poles of the Sun<br>- (b.2) No stereoscopic analysis and 3D reconstruction capabilities<br>- (c) No 3D localization of radio bursts<br>- (f) No high-resolution imaging of sunspots and CMEs<br>- (g) Does not provide in-situ measurements<br>- (h) Instruments have degraded drastically due to aging |
| HelioSwarm [16] | - (a.2) Incapability to study the poles of the Sun<br>- (c) No 3D localization of radio bursts<br>- (d) Limited and discontinuous forecasting of the arrival of CMEs and Solar Energetic Particles (SEP)<br>- (e) No continuous monitoring of the Sun-Earth Line (SEL)<br>- (g) Does not provide in-situ measurements |

Table 3. Existing Missions Analysis